\documentclass[prd,reprint,nofootinbib]{revtex4-1}
\usepackage{amsmath,amsfonts,graphicx,mathrsfs,subfig,psfrag,xcolor}

\newcommand{\re}[1]{(\ref{eq:#1})}

\def\phi{\varphi}

\def\rho{\varrho}
\def\d{\mathrm{d}}
\def\p{\partial}
\def\sgn{\mathop{\rm sgn}}

\renewcommand{\vec}[1]{\boldsymbol{#1}}

\newcommand{\rem}[1]{}

\def\={\discretionary{-}{-}{-}}

{\vskip 3pt plus 1pt minus 1pt\begin{normalsize}}%
{\par\end{normalsize}\vskip 3pt plus 1pt minus 1pt}

\newlength{\obrA} 
\newlength{\obrB} 
\newcommand{\cplx}[1]{\bar{#1}}
\newcommand{\ssp}{\scriptscriptstyle{+}}
\newcommand{\ssm}{\scriptscriptstyle{-}}

\newcommand{\Epsilon}{\mathcal{E}}
\newcommand{\fp}{\mathcal{A}}

\graphicspath{{./pictures/}{./}}

\begin{document}
\title{Rotating charged black holes accelerated by an electric field}
\date{\today}
\author{Ji\v{r}\'{i} Bi\v{c}\'{a}k}
\email{bicak@mbox.troja.mff.cuni.cz}
\author{David Kofro\v{n}}
\email{d.kofron@gmail.com}
\affiliation{
Institute of Theoretical Physics, Faculty of Mathematics and Physics,\\
Charles University in Prague,\\
V Hole\v{s}ovi\v{c}k\'{a}ch 2, 180\,00 Prague 8, Czech Republic}
\affiliation{Max Planck Institute for Gravitational Physics, Albert Einstein Institute,\\
Am M\"{u}hlenberg 1, D-14476 Golm, Germany}


\begin{abstract}
The Ernst method of removing nodal singularities from the charged C-metric representing uniformly accelerated black holes with mass $m$, charge $q$ and acceleration $A$ by ``adding'' an electric field $E$ is generalized. Utilizing the new form of the C-metric found recently, Ernst's simple ``equilibrium'' condition $mA=qE$ valid for small accelerations is generalized for arbitrary $A$. The nodal singularity is removed also in the case of accelerating and rotating charged black holes, and the corresponding equilibrium condition is determined.
\end{abstract}

\maketitle

\section{Introduction}
The only explicitly known solutions of the Einstein field equations describing moving finite objects are spacetimes with the boost and axial Killing vectors. They represent the exterior fields of ``uniformly accelerated particles'' which can be accelerated black holes like in the well-known C-metric \cite{kw}, or accelerated ``point singularities''  of various types like the Curzon-Chazy particles in the solution of Bonnor and Swaminarayan \cite{bszp}. For reviews of the boost-rotation symmetric spacetimes, including their radiative properties, see, e.g., \cite{bie}, \cite{pp-cjp} and references therein. Recently, we analyzed the Newtonian limit of these spacetimes using the rigorous Ehlers frame theory \cite{bikofnl}. The analysis corroborated their physical significance: the Newtonian limit describes the fields of classical point masses accelerated uniformly in classical mechanics.

In all cases other than those in which pairs of uniformly accelerated masses with one mass being negative occur, the particles do not move freely -- conical (``nodal'') singularities are present: They are interpreted as ``struts'' or ``strings'' necessary to produce the ``force'' accelerating the particles.

The conical singularities can be removed by introducing external fields.  By employing the Harrison-type transformation to the charged C-metric, Ernst \cite{ernst-rem} obtained the solutions to the Einstein-Maxwell equations describing a pair of oppositely charged black holes uniformly accelerated in a ``background'' electric field. If its strength characterized by parameter $E$ is properly chosen -- in the limit of the small acceleration parameter $A$ by the ``classical'' relation $qE=mA$ -- then the axis outside the black holes is regular.

\begin{figure}[]
\begin{center}
\includegraphics[keepaspectratio,width=0.4\textwidth]{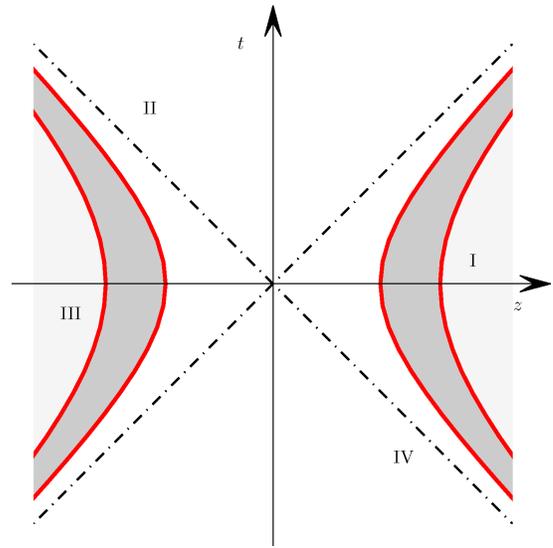}
\end{center}
\caption{The spacetime diagram of the C-metric in global coordinates.}
\label{fig:1}
\end{figure}




The external electric field is not compatible with asymptotical flatness. Independently, Bi\v{c}\'{a}k \cite{BiMo} calculated the motion of a charged black hole in an external electric field by means of the perturbative approach. These two results coincide to the degree of achieved precision. The advantage of the perturbative approach lies in the fact that the region filled with the appended electromagnetic field can be spatially bounded and thus the asymptotical flatness can be preserved. 

Ernst also removed the nodal singularity from the vacuum non-rotating C-metric by immersing it to external gravitational field \cite{Ernst-Gen}. The asymptotical flatness is, again, lost. 

In \cite{bihs-i} Ernst's procedure was applied to the Bonnor-Swaminarayan solutions and a new solution with two independent Curzon-Chazy particles falling freely in opposite directions in an external gravitational field was constructed. The same solution was then shown to follow from the Bonnor-Swaminarayan solution for two independent \emph{pairs} of accelerating particles if a limiting procedure is performed in which one particle in each pair is removed to infinity and its mass is simultaneously increased. This gives a clear physical interpretation also to the generalized C-metric solution of Ernst \cite{Ernst-Gen}.

More recently, Emparan \cite{Emparan} considered extremal black hole solutions with charges associated with various gauge fields of different ``stringy origin'' coupled to external field. The effects of such fields can compensate and a black hole can stay at rest, or one finds accelerating solutions of the Ernst type. Both static and accelerating non-extremal black holes coupled to different $U(1)$ gauge fields were also studied in \cite{Emparan}.

None of the previous work considered a free accelerating and simultaneously \emph{rotating} particle in an external field. In fact, no general theory similar to that given in \cite{BiSchm} is available for the boost-rotation symmetric spacetimes with Killing vectors which are not hypersurface orthogonal. However, there is one explicitly known metric available -- the spinning C-metric -- representing two charged, rotating black holes that are causally separated and accelerating in opposite directions.





The spinning/rotating C-metric has a fairly long history. Discovered by Pleba\'{n}ski and Demia\'{n}ski \cite{PD} as a subclass of their more general metrics (cf. \cite{GrPoBook}), later studied, for example, in \cite{FarZimm}, \cite{BiPr}, \cite{LeOl}, it was brought into a new form and slightly reinterpreted by Hong and Teo in 2005 \cite{HT2}. This new form has been inspired by bringing the ``structure function'' $\mathcal{G}(\xi)$ (see Eq. \re{rG} in Section \ref{sec:crCM}) into an explicitly factorizable expression which is not the case with the original form given in \cite{PD}. The first factorizable structure function for the non-rotating C-metric appeared, in fact, in the work of Emparan \cite{Emparan} mentioned above\footnote{Putting all charges $q_i$ in the formula (6.7) in \cite{Emparan} equal to zero we obtain the metric of an accelerating, uncharged and non-rotating black hole in the same form as later presented by Hong and Teo \cite{HT1} for the accelerating, non-rotating black hole. Even in the case with all $q_i=q$, Eq. (6.7) gives the metric in the factorizable form though not identical to that presented in \cite{HT1}.}. In this non-rotating case, the new form can be obtained by a coordinate transformation from the original one. It is not only more useful for calculations but also makes easier the interpretation of the parameters entering the C-metric.

In 2005 Hong and Teo \cite{HT2} turned to the \emph{rotating} C-metric. The ``traditional'' charged rotating C-metric contains regions near the rotation axis where the axial Killing vector is timelike, the causality in these regions is violated \cite{LeOl} and closed timelike curves can occur. This was interpreted as torsion singularities \cite{Le}. In \cite{HT2} it is shown that a new form with explicitly factorizable structure function exists in which conical singularities causing the acceleration of the rotating, charged black holes are such that no closed timelike curves -- no torsion singularities -- arise. Unlike with the non-rotating C-metric, the new form cannot be derived from the original one by a coordinate transformation. In \cite{BiKofAcc} we used this ``new'' C-metric to construct the field of an accelerated electromagnetic ``magic field'' in flat space (possibly desribing the classical model of a uniformly accelerated spinning electron); as a by-product  we have shown that this metric possesses a natural flat-space limit. In the present work we start out from this new form of the rotating C-metric.

The properties of the C-metric are rich even if the holes are non-rotating. The structure of conformal infinity of the whole class of the boost-rotation symmetric spacetimes is analyzed in \cite{BiSchm}. The analytic extensions through the horizons and the corresponding Carter-Penrose diagrams are constructed in \cite{GKP} by employing new form of the C-metric. In our Fig.\,\ref{fig:1} we plot only schematic spacetime diagram in the weak field limit of two -- possibly rotating -- black holes uniformly accelerated in opposite directions. (In the Weyl coordinates a black hole is represented by a rod, in the Fig.\,\ref{fig:1} we plot them as accelerated rods.)

In the present work we use new parametrization of the C-metric to generalize Ernst's results on removing nodal singularities by adding an external field to the case of arbitrary values of acceleration and to black holes which are rotating.

In the next Section \ref{sec:crCM} we review the basic properties of the charged rotating C-metric. The form of the metric of Hong and Teo \cite{HT2} is slightly modified by introducing a constant parameter scaling the azimutal coordinate. This makes the flat-space limit of the metric completely regular in the sense that no deficit angle around the symmetry axis arises in the limit. In Section \ref{sec:rem} the projection formalism and Harrison transformation are explained and the Ernst potentials for the charged rotating C-metric are found. Finally, in Section \ref{sec:rem2} the results are applied to discover ``the equilibrium condition'' which renders the rotating charged black hole falling freely in the external field. The original Ernst result $mA=qE$ valid for small acceleration $A$ and non-spinning black holes is generalized to the formula \re{remSm}. With just the first term in the expansion in gravitational constant $G$ included, this formula reads as follows:
\begin{equation}
mA= Eq\left[ 1+\frac{q^2A^2}{\left(1+a^2A^2\right)^2}\,G \right],
\label{eq:}
\end{equation}
where $a$ is the parameter characterizing the black hole rotation.

\section{The charged rotating C-metric} \label{sec:crCM}

The charged rotating C-metric\footnote{Thorough this paper we use the convention of the Exact Solutions book \cite{SC}, i.e., $G_{ab}=\kappa T_{ab}$, where $\kappa = 8\pi G c^{-4}$ and $T_{ab}= F_{ac}F_b^{\phantom{b}c}-\frac{1}{4}\,g_{ab}F_{cd}F^{cd}$.} in the slightly modified \cite{BiKofAcc} form of Hong and Teo \cite{HT2} reads
\begin{multline}
\d s^2 = \frac{1}{A^2(x-y)^2}\,\biggl\{ \frac{\mathcal{G}(y)}{1+\left( aAxy \right)^2}\;\Bigl[ \left( 1+a^2A^2x^2 \right)K\d t\\
 +aA\left( 1-x^2 \right)K\d\phi \Bigr]^2 -\frac{1+\left( aAxy \right)^2}{\mathcal{G}(y)}\,\d y^2 \\ 
+\frac{1+\left( aAxy \right)^2}{\mathcal{G}(x)}\,\d x^2 +\frac{\mathcal{G}(x)}{1+\left( aAxy \right)^2}\\
\times \Bigl[ \left( 1+a^2A^2y^2 \right)K\d\phi+aA\left(y^2-1\right)K\d t \Bigr]^2\biggr\}\,,
\label{eq:rCMr}
\end{multline}
where the structure function $\mathcal{G}(\xi)$ is 
\begin{equation}
\mathcal{G}(\xi) = \left( 1-\xi^2 \right)\left( 1+r_{\ssp}A\xi \right)\left( 1+r_{\ssm}A\xi \right)\,,
\label{eq:rG}
\end{equation}
with
\begin{equation}
r_\pm=G m \pm \sqrt{G^2m^2-a^2-G q^2}\,.
\label{eq:rrpm}
\end{equation}
Here $m$, $a$, $q$ and $A$ are, respectively, the mass, rotation, charge and acceleration parameter; we also keep the Newtonian constant $G$ in the metric which will turn out to be useful later. The metric \re{rCMr} slightly differs from that given in \cite{HT2} --- it is modified just by introduction of a constant $K=K(m,A,a,q,G)$  scaling the angular coordinate $\phi$, followed by a simple rotation\footnote{The same as in Eqs. (10), (11) in \cite{BiKofAcc}, i.e., $t\rightarrow K(1+a^2A^2)t$ and $\phi\rightarrow\phi-aAK^{-1}(1+a^2A^2)t$.}. It can be shown that if we put $G\rightarrow 0$ in \re{rCMr} we find the Minkowski space in accelerated spheroidal coordinates (see \cite{BiKofAcc}) but with a deficit angle in general. By choosing $K=(1+a^2A^2)^{-1}+O(G)$ we obtain the regular Minkowski spacetime in the limit $G\rightarrow 0$.

The structure function $\mathcal{G}(\xi)$ is constructed in such a way that  four simple real roots are
\begin{equation}
\begin{aligned} \xi_1 &= -\frac{1}{r_{\ssm}A}\,,\\ \xi_3 &= -1\,,\end{aligned} \qquad \begin{aligned}\xi_2 &= -\frac{1}{r_{\ssp}A}\,,\\ \xi_4 &= 1\,.\end{aligned}
\label{eq:realroots}
\end{equation}
(Compare with the original Demia\'{n}ski-Pleba\'{n}ski form \cite{PD} where $\mathcal{G}$ has very complicated roots.) The roots obey $\xi_1\leq\xi_2<\xi_3<\xi_4$. They determine several relevant regions: $x=y=\xi_3$ is infinity; the black hole event horizon corresponds to $y=\xi_2$; the acceleration horizon is at $y=\xi_3$; the line $x=\xi_4$ is the part of the symmetry axis extending from the black hole event horizon to the acceleration horizon and $x=\xi_3$ connects the event horizon with infinity. The roots determine the allowed ranges of $x$ and $y$ coordinates in \re{rG} as follows: $y\in \langle \xi_2,\, \xi_3 \rangle$ for quadrant I (see Fig. \ref{fig:1}), $y\in\langle \xi_3,\,\xi_4\rangle$ for quadrant II and $x  \in \langle \xi_3,\,\xi_4 \rangle$.

The black hole is located in the quadrant I where the boost Killing vector field is timelike. Nevertheless, the metric \re{rCMr} describes also the quadrant II where no black holes are present. This regime is dynamical and radiative (the boost Killing vector is spacelike).

The 4-potential of the electromagnetic field is
\begin{equation}
\vec{\fp} = \frac{Kqy\ \left[ \left(1+a^2A^2x^2\right)\vec{\d} t+aA\left( 1-x^2 \right)\vec{\d}\phi \right]}{\sqrt{4\pi}\left[1+\left( aAxy \right)^2\right]}\,.
\label{eq:rA}
\end{equation}

The axial Killing vector field is $\vec{\xi}=\vec{\p_\phi}$; let us denote $F=\xi^a\xi_a$. The axis of symmetry is regular if at the axis $F^{,a}F_{,a}/4F \rightarrow 1$ (see, e.g., \cite{SC}, Eq. (19.3)). Calculating this invariant for the C-metric we get
\begin{equation}
\delta = \frac{F^{,a}F_{,a}}{4F} = \frac{A^2\left( x-y \right)^2}{1+\left( aAxy \right)^2} \frac{\bigl( \mathcal{G}(x)F^2_{,x}-\mathcal{G}(y)F^2_{,y} \bigr)}{4F}\,.
\label{eq:}
\end{equation}
The parameterization of the C-metric is chosen such that the axis is given by $x=-1$ or $x=1$. The regularity condition for the metric \re{rCMr} becomes
\begin{equation}
\lim_{x\rightarrow \pm1} \delta = K^2 \left[ 1+a^2A^2 +G\left( A^2q^2\pm 2Am \right)\right]^2 = 1\,.
\label{eq:rDefAo} 
\end{equation}
Clearly, this is not valid in general; it can be satisfied either at $x=+1$ or at $x=-1$ by a suitable choice of $K$; but it cannot be satisfied at both parts of the axis $x=\pm 1$. (An accelerating black hole must be attached to a string at ``one side'' at least.)

The net physical charge of the whole spacetime is zero, the physical charge $Q$ of one of the black holes is
\begin{equation}
Q = \frac{1}{\sqrt{4\pi}} \int_{S} \star \vec{F} = K q
\label{eq:charge}
\end{equation}
(here $S$ is any closed 2-surface surrounding the black hole).
The total angular momentum is zero as the black holes are counterrotating. The total angular momentum of one of the black holes is simple in the vacuum case:
\begin{equation}
J = -\frac{1}{8\pi}\,\int_{S} \star \,\vec{d\xi_{(\phi)}} = K^2Gam\,.
\label{eq:am}
\end{equation}
Possible contributions due to the charge are of order $O(q^2)$. From \re{charge} and \re{am} it can be seen that parameter $K$ enters physical quantities.
 
\section{Removal of the nodal singularity of the charged rotating C-metric}\label{sec:rem}
In order to remove nodal singularities we shall employ the Harrison transformation in which the quantities entering so-called projection formalism \cite{SC,Geroch,Racz} and the Ernst equations \cite{Ernst-AxI,Ernst-AxII} for complex potentials $\Epsilon$ and $\Phi$ appear. The method employs the existence of the Killing vector field $\vec{\xi}$ over a spacetime manifold; this gives rise to a uniquely defined quotient 3-manifold.

\subsection{Projection formalism and the Harrison transformation}\label{sec:ProjF}
Let there exist the Killing vector field $\vec{\xi}$. The projection is usually made with respect to a timelike Killing vector but the formalism is easily modified to a spacelike one. In Sec. \ref{sec:rem2} we specialize to the axial Killing vector $\vec{\xi_{(\phi)}}$. 

Define the scalar field $F$ and the twist vector $\vec{\omega}$ by
\begin{equation}
F = \xi^a\xi_a\,,\quad \epsilon=\sgn F\,,\quad \omega_a = -\epsilon_{abcd}\xi^b\nabla^c\xi^d\,.
\label{eq:PFfodef}
\end{equation}
Then the metric induced on the quotient manifold after the conformal transformation (given by $F$) is (see, e.g., \cite{SC})
\begin{equation}
\gamma_{ab} = \epsilon F \left(g_{ab}-\xi_a\xi_b/F\right).
\label{eq:PFindM}
\end{equation}
The Einstein-Maxwell system can be reduced to Ernst equations 
\begin{subequations}
\begin{eqnarray}
-F\,\tilde{\square}\, \Epsilon &=& \left( \tilde{\nabla}_{\! a}\Epsilon+2\,\cplx{\Phi}\,\tilde{\nabla}_{\! a}\Phi \right)\tilde{\nabla}^a\Epsilon\,, \\
-F\,\tilde{\square}\, \Phi &=& \left( \tilde{\nabla}_{\! a}\Epsilon+2\,\cplx{\Phi}\,\tilde{\nabla}_{\! a}\Phi \right)\tilde{\nabla}^a\Phi\,,	
\end{eqnarray}
\label{eq:ErnstEQ}
\end{subequations}
where $\tilde{\nabla}$ is the covariant derivative associated with $\gamma_{ab}$ and $\tilde{\square} = \gamma^{ab}\tilde{\nabla}_{\! a}\tilde{\nabla}_{b}$.
The electromagnetic Ernst potential $\Phi$ is defined by
\begin{equation}
\frac{\sqrt{2\kappa}}{2}\;\xi^aF^*_{ab}=\Phi_{,b}\,,\qquad \Phi_{,a}\xi^a=0\,,\qquad F^{*ab}_{\phantom{*ab};b}=0\,,
\label{eq:ErnstPhi}
\end{equation}
and the gravitational Ernst potential $\Epsilon$ is given by 
\begin{eqnarray}
\Epsilon_{,a} &=& -F_{,a}+i\omega_a-2\,\cplx{\Phi}\Phi_{,a}\,,\\
-F &=& \frac{1}{2}\left( \Epsilon+\cplx{\Epsilon} \right) + \cplx{\Phi}\Phi\,.
\end{eqnarray}
The existence of $\Epsilon$ is guaranteed by (see \cite{SC})
\begin{subequations}
\begin{align}
\xi^aK^*_{ab}&=\Epsilon_{,b}\,,& \Epsilon_{,a}\xi^a&=0\,,\\
K^{*ab}_{\phantom{*ab};b}&=0\,, &K^*_{ab} &= -2\xi^*_{a;b}-\sqrt{2\kappa}\,\cplx{\Phi}F^*_{ab}\,,
\label{eq:ErnstEpsilon}
\end{align}
\end{subequations}
where $X_{ab}^* = X_{ab} + \frac{i}{2}\,\epsilon_{abcd}X^{cd}$ is the complex self-dual bivector.

The form of the Ernst equations is preserved under 8-parameter group $G_8$ of transformations of the potentials. We shall employ just the Harrison transformation belonging to $G_8$. Introducing $\Lambda$ by
\begin{equation}
\Lambda = 1-2\cplx{\gamma}\Phi-\cplx{\gamma}\gamma\Epsilon\,,
\label{eq:Lambda}
\end{equation}
the Harrison transformation reads 
\begin{equation}
\hat{\Epsilon} = \Epsilon \Lambda^{-1}\,,\qquad \hat{\Phi} = \left( \Phi+\gamma\Epsilon \right)\Lambda^{-1}\,.
\label{eq:Harrison}
\end{equation}
The norm of the Killing vector transforms as
\begin{equation}
\hat{F} = -\frac{\frac{1}{2}\left( \Epsilon+\cplx{\Epsilon} \right)+\Phi\cplx{\Phi}}{\cplx{\Lambda}\Lambda}= \frac{F}{\cplx{\Lambda}\Lambda}\,.
\label{eq:trfKilling}
\end{equation}
The new metric can be reconstructed in the form $\hat{g}_{ab} = |\hat{F}^{-1}|\,\gamma_{ab}+\hat{F}^{-1}\hat{\xi}_a\hat{\xi}_b$ following \cite{SC}, nevertheless, it will not be needed to investigate the properties of the axis\footnote{To evaluate $\hat{F}^{,a}\hat{F}_{,a}/\hat{F}$ in our coordinate system, we need to know just the metric on surfaces of constant $\phi$ and $t$ because function $\hat{F}$ depends on $x$ and $y$ only, i.e., $\hat{F}=\hat{F}(x,y)$. The necessary metric components can be read off directly from the conformal metric.}.

\subsection{Ernst potentials for the charged rotating C-metric}\label{sec:Ernst}
To generate a singularity-free solution from the charged, rotating C-metric \re{rCMr} with the help of the Harrison transformation \re{Harrison} we need to know its generating complex potentials associated with the \emph{axial} Killing vector field $\vec{\xi_{(\phi)}}$.

We know the full solution and thus we can find the norm of $\vec{\xi_{(\phi)}}$ explicitly 
\begin{equation}
F = K^2\,\frac{a^2A^2\left( 1-x^2 \right)\mathcal{G}(y)+\left( 1+a^2A^2y^2 \right)\mathcal{G}(x)}{A^2\left( x-y \right)^2\left[ 1 +\left(aAxy  \right)^2\right]}\,.
\label{eq:CMF}
\end{equation}
The complex electromagnetic potential $\Phi$ is follows from the integration of \re{ErnstPhi} in which the field $F^{ab}$ can be determined from the 4-potential \re{rA}. Rather lengthy calculations lead to a surprisingly simple result:
\begin{equation}
\Phi =  -\sqrt{G}\,Kq\frac{aAy-ix}{1+iaAxy}\,.
\label{eq:cplxPhi}
\end{equation}
The gravitational Ernst potential $\Epsilon$ can be obtained by integrating \re{ErnstEpsilon}. Its form is more complicated:
\begin{multline}
\Epsilon = -F - \cplx{\Phi}\Phi +i2aK^2G\,{(x-y)^{-1}\left( 1+\left( aAxy \right)^2 \right)^{-1}}\\ 
\Bigl[ yA(x^2-a)(1-xy) + ya^2m(1-3x^2+yx^3+yx)\\
 -m(x^2+1-3xy+yx^3\Bigr]\,,
\label{eq:CME}
\end{multline}
where $F$ is given by \re{CMF} and $\Phi$ by \re{cplxPhi}. 

The potential $\Epsilon$ can also be written in the form analogous to the potential $\Phi$, 
$$\Epsilon = \frac{P(x,y)}{A^2\left( x-y \right)^2\left( 1+iaAxy \right)}\,,$$
but we have not yet found a compact expression for the complex polynomial $P(x,y)$. However, it is merely a polynomial of order 3 in variables $x,\,y$.

\section{Removal of the nodal singularity of the charged C-metric: rotating case}\label{sec:rem2}
With $\Epsilon$ and $\Phi$ known we employ the transformation \re{Harrison}. Let us choose parameter $\gamma$ as  
\begin{equation}
\gamma = -\frac{1}{2}\,i\sqrt{G}\,E\,.
\label{eq:}
\end{equation}
In fact, up to the factor $\sqrt{G}$ this form of $\gamma$ follows that of \cite{Ernst-Gen} but the parameter leading to the regular axis will now be different. In the special-relativistic limit the parameter $E$ can be interpreted as the strength of the (added) external homogeneous electromagnetic field.

The norm of the Killing vector transforms as in \re{trfKilling}, i.e., $\hat{F} = {F}/{\cplx{\Lambda}\Lambda}$ where $F$ follows from \re{CMF} and $\Lambda$ from \re{Lambda} with $\Phi$ and $\Epsilon$ given by \re{cplxPhi} and \re{CME}. The invariant to be evaluated at the axis turns out to be 
\begin{equation}
\hat{\delta}=\frac{\hat{F}_{,a}\hat{F}^{,a}}{4\hat{F}} = \frac{A^2\left( x-y \right)^2}{1+\left( aAxy \right)^2} \frac{\Bigl[\, \mathcal{G}(x)\hat{F}^2_{,x}-\mathcal{G}(y)\hat{F}^2_{,y}\,\Bigr]}{4F}\,.
\label{eq:}
\end{equation}
Evaluating the limits at the symmetry axis, i.e. at $x\rightarrow \pm 1$, leads to
\begin{equation}
\lim_{x\rightarrow \pm 1}\hat{\delta} = \frac{K^2\left( 1+a^2A^2+G\left( q^2A^2\pm 2mA \right) \right)^2}{\left[\left(1\pm \frac{1}{2}\,G KEq\right)^4+G^4K^4E^4a^2m^2\right]^2}\,. \label{eq:CMrp} 
\end{equation}

The ``equilibrium'' conditions guaranteeing the equality of conical singularities from both sides read:
\begin{equation}
\lim_{x\rightarrow 1}\hat{\delta}=\lim_{x\rightarrow -1}\hat{\delta} = 1\,,
\label{eq:eqc}
\end{equation}
which is a system of two equations for two unknown parameters, say, $K$ and $m$. This system can be explicitly solved but the final result is formidable. However, when expanded in $G$ the result becomes nicely simple and intuitive. Denoting 
\begin{equation}
\alpha = 1+a^2A^2\,,
\label{eq:alpha}
\end{equation}
we get 
\begin{subequations}
\begin{eqnarray}
\alpha K &=& 1-\frac{q^2A^2}{\alpha}\, G  + \frac{\left( 2q^2A^4+3E^2 \right)q^2}{2\alpha^2}\, G^2 + \ldots\,,\ \ \ \ \ \  \label{eq:remSK} \\
m &=& \frac{qE}{A}\left[ 1+\frac{E^2q^2}{4\alpha^2}\,G^2 -\frac{E^2q^4A^2}{2\alpha^3}\,G^3+\ldots \right]\!.\ \ \ \ \ \ \label{eq:remSm}
\end{eqnarray}
\label{eq:sol}
\end{subequations}

Equation \re{remSm} resembles the classical relation for motion of a particle with charge $q$ in electric field $E$ plus $G$-dependent corrections. The interpretation is modified by the fact that the net physical charge is given by \re{charge} -- $Q=Kq$ -- and thus the last equation in terms of the net charge $Q$ reads
\begin{multline}
\frac{m}{\alpha} = \frac{EQ}{A} \Biggl[ 1 + \frac{1}{4}\,E^2Q^2G^2 \\
-\frac{1}{2}\alpha E^2Q^4A^2G^3 + \ldots \Biggr]\,.
\label{eq:remSm-massR}
\end{multline}
Notice that for small accelerations $|mA|\ll 1$, factor $K\simeq 1+O\left( (aA)^2 \right)$, so that \re{remSm} and \re{remSm-massR} coincide in this limit. 

Equation \re{remSm-massR} indicates what appears to play the role of mass, namely $M=Km$. It should be emphasized that our source becomes point-like in the limit $G\rightarrow 0$ only if simultaneously $a\rightarrow 0$. Then $A$ is acceleration with respect to the flat background. For $a\neq 0$, the source is a rotating disc with complicated structure (see \cite{BiKofAcc} for details).

\begin{acknowledgments}
We acknowledge the support from the Grants No. MSM0021620860 of the Ministry of Education of the Czech Republic, No. LC06014 (``Centre of Theoretical Astrophysics'') and No. GA\v{C}R 202/06/0041. D.K. also acknowledges the partial support from the Grant No. GAUK 116-10/258025. We are also grateful to the Albert Einstein Institute, Golm, for the kind hospitality.

\end{acknowledgments}

\end{document}